# Stability of Granular Tunnel

Elfi Yuliza, Nadya Amalia, Handika Dany Rahmayanti, Rahmawati Munir,

Muhammad Miftahul Munir, Khairurrijal Khairurrijal, Mikrajuddin Abdullah[a]

Department of Physics, Institut Teknologi Bandung

Jalan. Ganesa 10, Bandung 40132, Indonesia

[a]Email: mikrajuddin@gmail.com

**ABTRACT**

We demonstrated the stability of tunnels made of granular matters is strongly dependent on the grain size, tunnel diameter, and water content in the granules. Larger tunnel radius, larger grain size, and too much water content tend to destabilize the tunnel. We also develop a model to describe such findings. We indentified a phase diagram of stability which greatly controlled by granular bond order. For granular bond order of larger than unity, we can alwaysmade a stable tunnel. However, for granular bond order of less than unity, we obtain a general expression for maximum tunnel thickness that can be made. To best of our knowledge, this is the first exploration regarding the granular tunnel stability.

**INTRODUCTION**

There are so many interesting physical phenomena exhibited by granular matters and there are so many properties of granular matters that have been explored and reported, covering from natural phenomena to industrial processes. Suchproperties consist of both static (granules are resting) and dynamic (granules are moving)situations. The static behaviors are commonly investigated in an attempt to explain the stability of "granular building". The investigated parameters are very diverse such as grain size, shapes, density, surface, wet level of the surface, etc. It is clearly proven that those parameters directly or indirectly influce the stability of the "granular building". For example, the size of sandcastlesis strongly affected by grain size, shape and water content inside.One of the most spectacular and fascinating properties is how the addition of a small amount of fluid dramatically changes the macroscopic properties of the material [1]. The liquid mixed with the granules will generate cohesive forces between grains which profoundly influence the macroscopic stability of granular piles[2]. Just a bit of water turns a boring pile of dry sand into a spectacular sandcastle [3-5] while too much water will destabilize the material, the mechanism that migh causes landslide disasters [6].

The relation between the physical properties of the grain and the dimension of sandpiles or sandcastles are prominently chalenging and clearly showing how small-scale surface properties dramatically influnece large-scale mechanism [7]. More importantly, understanding these relationship can provide key insight into various phenomena in nature, especially those related to soil or grain displacement[2]. For some landslides, the basal material is more cohesive than the flowing one, a situation arises for example because of humidity [8] which is directly related to water content in the granules. Furthermore the properties of granular materials are of huge importance to engineers and it is estimated that about 10% of all energy consumption on Earth is spent on the handling of granular materials [9].

In this work we will investigate the static behavor of granular matters, especially the stability of "granular building". Indeed, there are limited "granular building" have been well explored, to date, the more popular ones are sandcastle, granular piles, and "granular mountain". We are interested in exploring one different "granular building", namely granular tunnel, i.e. a tunnel made at the bottom of granular material as shown in **Fig. 1**. To best of our knowledge, no report has been published discussing to this topic. We will do experiments to identify what parameters that are responsible for controlling the stability of the tunnel and then develop a model to explain the observation.

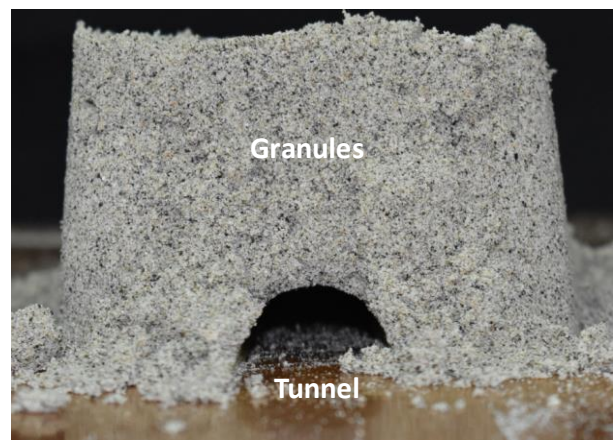

**Figure 1.** An example picture of a tunnel. A tunnel of half cylindrical shape is created at the bottom of granules.

**EXPERIMENT**

The tunnels were made by putting half cylinder of PVC tube inside a topless plastic box. The box dimension is large enough compared to tunnel radius to guarantee the edge effect is negligble. The granules were poured gently from the open top of the box and then pressed at the same pressure. The box was then pulled upwars to leave the granules unsupported from all sides. Finally, the half cylinder tube was pulled gently to leave granular tunnel.

**RESULTS**

First, we investigated the stability of tunnel using different grain sizes and different tunnels diameters. We made tunnels of the same height, 14 cm (measured from the top of the tunnel to the top flat granular surface). **Figure2(A)** informs the stability of the tunnels. All tunnels have been made using granules containing a water content of 2.43 % w/w. The stable condition means the tunnel can be made, while the unstable condition means the tunnel suddenly abrupted when removing the half cylinder support. Using grain of diameters 0.1 mm and 0.5 mm we were able to make tunnuels at radius of up to 3.2 cm. But when using granular of diameter 1.5 mm, we failed to make tunnels of diameters 3.2 cmand larger. This is consistent with some reports mentioning that fine particles are strongly agglomerated [10], and the ability of making tunnels means the particles are strongly agglomerated (strongly attract each others).

**Figure 2(B)** displays the effect of water content on the tunnel stability. Seven water contents were investigated: 0 %w/w (dry granules), 2.43 %w/w, 4.76 %w/w, 6.98 %w/w, 9.1 %w/w, 13 %w/w, and 14.9 %w/w. We used the same granules of grain diameter0.5 mm and the tunnel heights were maintained at 14 cm. We were unable to make tunnels using dry granules or granules containing water contents of 14.9 %w/w and above, especially in the experimental condition explored here.It is consistent with some reports, mentioning that when the distance between particles reaches a certain value, the liquid bridge is damaged, and the capillary attractions between particles disappear[11,12]. When increasing the liquid fraction, the distance between granules increases, or too much water will destabilize the "granular building" [6]. Since we used natural sands, the surfacesare not smooth. For granular having rough surface, at very low water content, most of the water is trapped in the surface roughness, and the bridge force is dominated not by the curvature of the sphere, but by the local roughness. At higher water content, a significant fraction of the water is still caught in the surface roughness but the bridge force is dominated by the curvature of the spheres. At even higher water contents, the bridges start merging and loose strength [1].

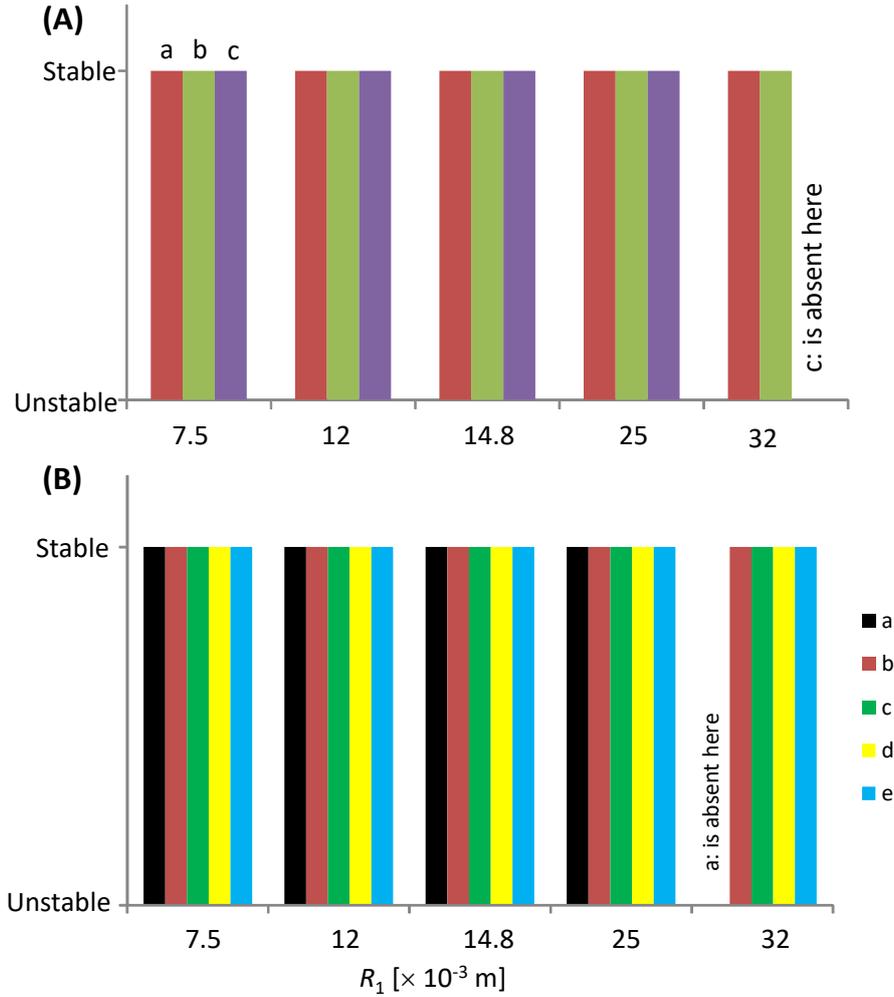

**Figure 2**. (A) Stability diagram of tunnel up to 14 cm height as function of tunnel radius. We take data for varius grain diameters: (a) d = 0.1 mm, (b) d = 0.5 mm, and c) d = 1.5 mm. The bar height only indentifies the stability: bar exists to mean stable and bar is absent to mean unstable (fail to create the tunnel). (B) Stability diagram of tunnel up to 14 cm theight as function of tunnel radius. We used granules of grain diameter 0.5 mm. We take data for various water contents (% w/w): (a) 2.43, (b) 4.76, c) 6.98, (d) 9.1, and (e) 13%. For water contents of 0% (dry) and 14.9% and above we failed to create stable tunnels.

## MODELING AND DISCUSSION

In this model we focused on explaining how the grain size, tunnel radius, and water content control the tunnel stability. The tunnel geometry was a half of cylinder havingradius *R* and infinity length as illustrated in **Figure 3(left)**. The boundary effect is neglected by assumingthat all edges are located far away. The granules, especially those located near the tunnel curvature, arrange according to a circular symmetry centering at the tube axis. We assume the grain sizes are all identical so that the thickness of the tunnel (measured from the top of the tunnel to the top flat granular surface) is $h = Md$ with *M* is the number of layer above the tunnel peak. The lower most layer which in contact with the tunnel curvatureis assigned as the 1st layer and the uppermost layer as the $M^{th}$ layer.

To explain the tunnel stability, let us focus on a central column located just above the tunnel peak, made by stack single particles of height $h$. The external forces acting on each particle is illustrated in **Figure 3(right)**: supporting force by nearest neighbor particles, cohesive force, and weight. Similar consideration has also been discussed by Nowak et al when discussing the stability of wet granular pile [5].

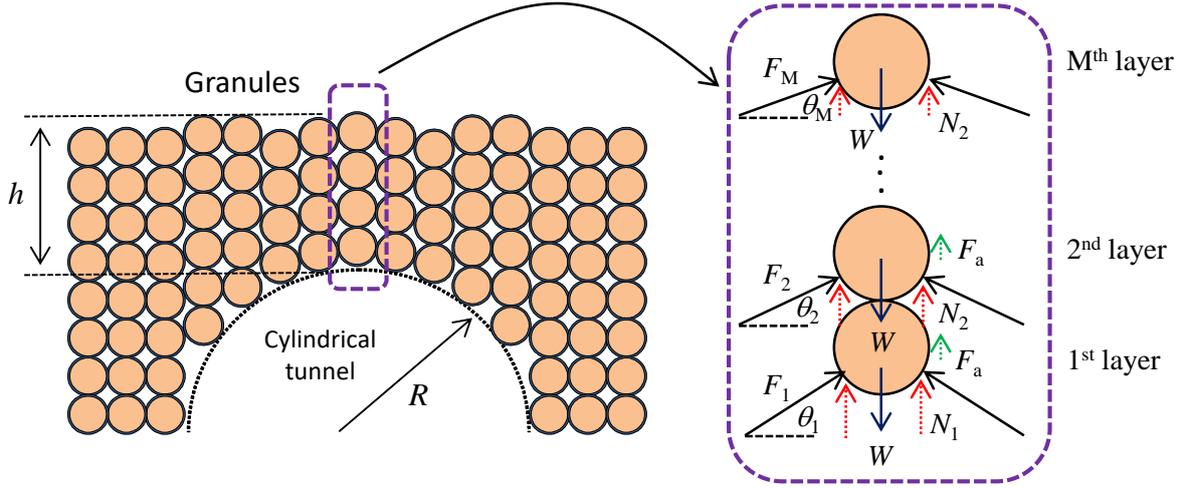

**Figure 3** (left) Ilustration of a granular tunnel. (right) One column developing the center of the tunnel as a stact of $M$ grains. It is also illustrated three external forces acting of the grain column: supporting forces ($F_i$), capillary forces ($F_a$), and the grain weight ($W$).

Suppose the number of supporting particles below each particle in the column, excluding by particles forming the column, is $z$. We excluded the supporting force by other particles in the column since we are interested only in external forces acting to the column to apply the Newton law. Each supporting particle at the i-th layer produces a force $F_i$ making an angle $\theta_i$ to the horizontal. The vertical component of this force is $N_i = F_i \cos\theta_i$, and the total vertically supporting force experienced by the i-th particle in the column is $zN_i$. Therefore, the total supporting force experienced by all particles in the column is

$$F_{\sup} = \sum_{i=1}^{M-1} zF_i \sin\theta_i \qquad (1)$$

Each particle experiences a capillary force by fluid that is trapped in space between particles. Becase the number of layer is $M$, the total number of space filled by fluid (space between layers) is $M$-1. By assuming the fluid is distributed homogeneously throughout the granules, the uprward capillary forces experiences by parcles in the 1st layer to the $M$-1th layer are the same so that the total upward capillary forces experienced by the column is $F_{cap} = MF_a$, with $F_a$ is the upward cappilary forces acting to a particle. The capillary force (force produced by a liquid bridge) can be approximated as $F_a = (\alpha\pi/2)\Gamma\tilde{d}$, with $\alpha$ is a dimensionless constant, depending on the dimension and shape of the liquid bridge, $\Gamma$ is the liquid surface tension, and $\tilde{d}$ is the diameter of curvature (2 × radius of curvature) of the liquid bridge between particles [12,13]. The order of magnitude of $\tilde{d}$ is nearly the same as that of a grain, so that for simplicity we can assume $\tilde{d} \approx d$. Possible sligtly deviation in the radius for curvature of the liquid bridge byequalizing it with the grainradius might becorrected by sligthly adjusting the value of parameter α. In several models such as Weigert's model [14], Willet's full model and Willet's reduced model [15], and Rabinovich's model [16], the diameter used for calculating the cohesive force is the effective diameter of the contacting grains, $d_{ef} = 2d_i d_j/(d_i + d_j)$ [17] which becomes equal to grain diameter when the diametersof all grainss are identical. The number of bridges has been demonstrated to increase abruptly with volume fraction (from about one per sphere to about six) at a volume fraction of about 0.2% [18]. However, this possible change in the liquid bridge number is not included in our model as also treated by Møller and Bonn [1].

The total supporting and capillary forces must be able to overcome the column weight to ensure the tunnel stability, or $F_{sup} + F_{cap} - MW = 0$ with $W = mg$ is the weight of a granular. Similarly, Nowak et al have discussed the stabiliy of a granular supported by some grains from the bottom and how the support force and the cappilary force contribute to the stability to overcome gravitation [5].

Let us temporarily define $\alpha_i = F_i/F_1$ so that Eq. (1) can be rewritten as $F_{sup} = MzF_1\left(\sum_{i=1}^{M-1}\alpha_i \sin\theta_i / M\right)$ and the condition for stability is expressed as

$$zF_1\left(\frac{1}{M}\sum_{i=1}^{M-1}\alpha_i \sin\theta_i\right) + F_a - W = 0 \tag{2}$$

By considering **Fig. 3(right)**, it is clear that $F_1 > F_i$, i=2,3,... so thatwe conclude that $0 < \alpha_i \leq 1$.

We will simplify the summation by writting $\sin\theta_i = d/R_i$. We denote the radius of the first layer above the tunnel void (the tunnel curvature) as $R_1$. The radius of the i-th layer is $R_i = R_1 + (i-1)d$ so that $\sin\theta_i = d/(R_1 + (i-1)d) = 1/((x-1)+i)$, with $x = R_1/d$. We also see from **Fig. 3(right)** that the supporting granules in the 1st layer supports $M$ grains (entire particles in the column), the supporting particles in the 2nd layers support $M$-1 grains, etc. Thus we can roughly approximate $F_i \propto M-(i-1)$, resulting $\alpha_i \approx [M-(i-1)]/M = (1+1/M) - i/M$ and

$$zF_1\left(\frac{1}{M}\sum_{i=1}^{M-1}\alpha_i \sin\theta_i\right) \approx zF_1\left(\frac{1}{M}+\frac{x}{M^2}\right)[\psi(0,M-1+x)-\psi(0,x)]-zF_1\frac{M-1}{M^2} \qquad (3)$$

with $\psi(0,z)$ is the digamma function, defined as $\psi(0,z)=d\ln\Gamma(z)/dz$, with $\Gamma(z)$ is the gamma function.

If we assume that the number of layer is much larger than unity we may neglect the last term in Eq. (3) so that we may approximate $zF_1\left(\sum_{i=1}^{M-1}\alpha_i \sin\theta_i/M\right) \approx zF_1 H(x,M)$ with a definition

$$H(x,M)=\left(1+\frac{x}{M}\right)\frac{[\psi(0,M-1+x)-\psi(0,x)]}{M} \qquad (4)$$

The stable condition in Eq. (2) is then approximated as $zF_1 H(x,M)+F_a-W=0$ or

$$\frac{zF_1}{W}H(x,M)+Bo_g-1=0 \qquad (5)$$

with $Bo_g = F_a/W$ is the cohesive granular bond order [10]. For large values of this number the grains tend to aggregate.

As an illustration we will analyze the stability of a tunnel made of very fine grains of diameter diameter 0.1 mm. We used various tunnel radii: 7.5 mm, 12 mm, dan 24 mm, which correspond to $x$ = 75, 120, dan 240, respectively. **Figure 4(A)** is a plot $H(x,M)$ with respect to $M$ for three above parameters. It is obvious that as $M$ increases, $H(x,M)$ decreases. It is also clear from **Figure 4(A)** that $H(x,M)$ is always positive.

**Figure 4(B)** is a plot of $zF_1 H(x,M)+Bo_g-1$ as function of $M$. In calculation we used $x$ = 100 and $zF_1/W$ = 50 as an example. The positive values correspond to stable state and the negative values correspond to unstable states. It is clear that the number of layer to ensure the stable state increases with the $Bo_g$. For $Bo_g> 1$ all values are positive to mean that the tunnels can be made at any layer thickness. However, for $Bo_g< 1$, there is a maximum $M$ that can produce the stable state, and the maxuimum $M$ decreases as the $Bo_g$ decreases below unity.

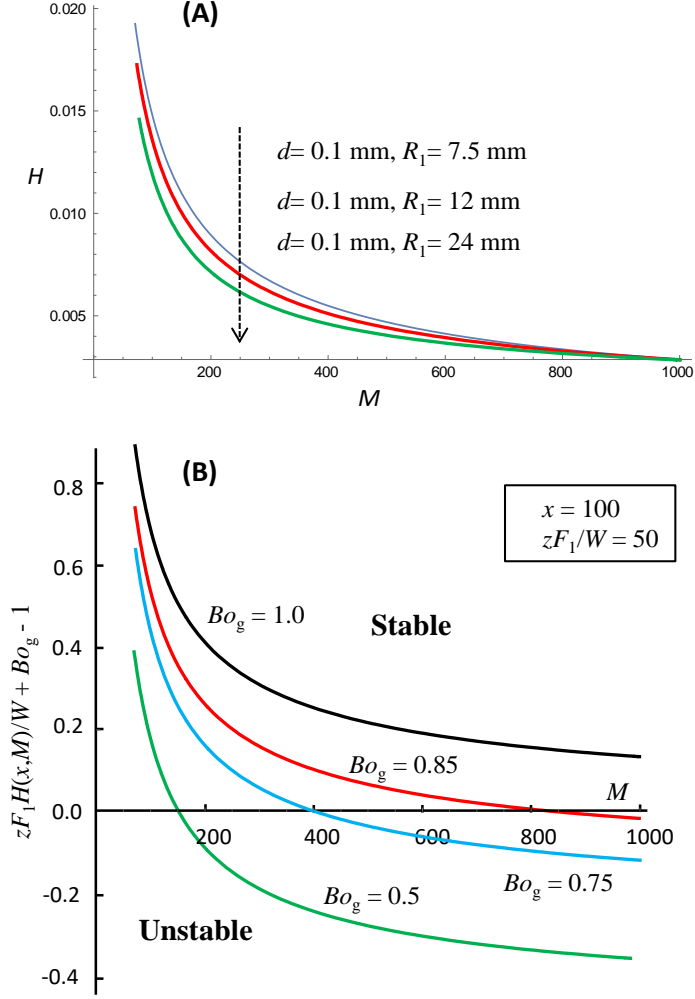

**Figure 4**(A) Dependency of $H(x,M)$ on $M$ for different $x = R_1/d$. The graindiameter is fixed at 0.1 mm, while the tunnel radius is varied. The top curve, middle curve, and bottom curves correspond to $R_1$ = 7.5 mm, 12 mm, and 24 mm, respectively. (B) Dependency of $zF_1H(x,M)/W + Bo_g - 1$ on $M$ for different $Bo_g$. The ratio of $x = R_1/d$ was fixed at 100 and $zF_1/W$ was fixed at 50. The curves from the top to the bottom correspond for $Bo_g$ = 1.0, 0.85, 0.75, and 0.5, respectively.

Now let us analyze the boundary for $H(x,M)$. Let us look again the derivation of Eq. (3) and inspect the following term

$$\left(\frac{1}{M}+\frac{x}{M^2}\right)\sum_{i=1}^{M-1}\frac{1}{(x-1)+i} = \frac{1}{M}\sum_{i=1}^{M-1}\left(1+\frac{x}{M}\right)\left(\frac{1}{(x-1)+i}\right) \quad (6)$$

The largest term in the summation is the first term, occuring when $i = 1$, i.e., $(1+x/M)(1/x) = 1/x+1/M$. Because the problem we are investigating is tunnels having radii of much larger than the grain diameter we will always have $x > 2$, and the number of layers is much larger than 2 so that we will always have $M > 2$ then $1/x + 1/M < 1/2 + 1/2$, $1/x + 1/M < 1$. With this criterion, Eq. (6) can be rewritten as the following inequality $(1/M)\sum_{i=1}^{M-1}(1+x/M)/((x-1)+i) < (1/M)\sum_{i=1}^{M-1}1 = (M-1)/M < 1$. Thus we conclude that $H(x,M) < 1$.

Since there is a maximum value of $F_1$, i.e., $F_{1,\max}$, then to ensure that the tunnel does not abrupt, the stable state satisfies $zF_{1,\max}H(x,M) + F_a - W > 0$, or

$$\frac{zF_{1,\max}}{W}H(x,M) + Bo_g - 1 > 0 \tag{7}$$

The first term on the left hand side of Eq. (7) is always positive but decreases when $M$ increases. Therefore if $Bo_g > 1$, the left term is always positive tomean the tunnel will stable at any $M$ or at any altututdesif $Bo_g > 1$. Even, the stable tunnel can be made at the condition of $F_{1,\max} = 0$ as long as $Bo_g > 1$.

To the contrary, if $Bo_g < 1$, the stability is strongly dependent on $F_{1,\max}$ at a certain tunnel thickess. The stable and unstable state is separated by a line

$$F_{1,\max} = (1 - Bo_g)\frac{W}{zH(x,M)} \tag{8}$$

**Figure 5** shows the phase diagram of the tunnel stability. For $Bo_g > 1$, all regions are stable state and for $Bo_g < 1$, only states above the curve are stable and states below the curve are unstable.

$Bo_g = 1$ is considered to be the transition between the cohesive and adhesive region [10]. There is a maximum $M$, below which the tunnel is stable and above which the tunnle is unstable. The maximum $M$ satisfies $(zF_{1,\max}/W)H(x,M_{\max}) - |Bo_g - 1| = 0$. A special case if $Bo_g = 0$ (no cohesive force between granules), then the tunnels stability will be achieved when $(zF_{1,\max}/W)H(x,M) > 1$ or

$$F_{1,\max} > \frac{W}{zH(x,M)} \tag{9}$$

for any finite $M$.

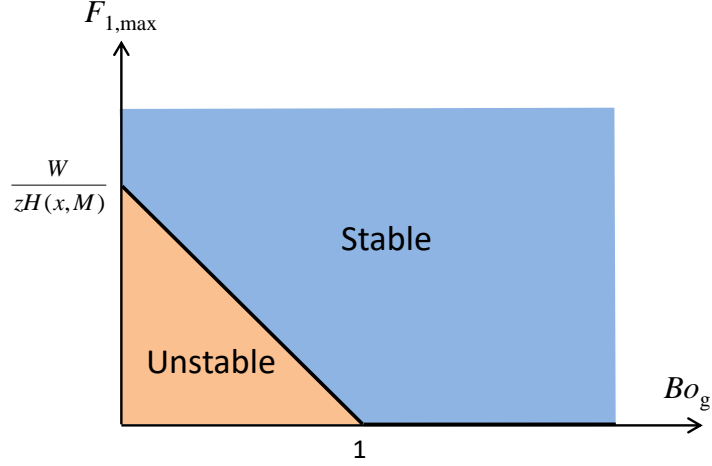

**Figure 5** Phase diagram separating the stable and unstable state.

Inded, the $Bo_g$ depends on the grain diameter. The explisit expression for the $Bo_g$ is

$$Bo_g = \frac{3\alpha\Gamma}{d^2\rho g} \tag{10}$$

Thus the condition for stability can be rewritten as $(zF_{1,\max}/W)H(x,M)+(3\alpha\Gamma/\rho g d^2)-1>0$. As an illustration, water has $\Gamma \approx 70$ dyne/cm $= 0.07$ N/m [5]. The parameter $\alpha$ has an order of unity so that the condition for stability is $(zF_{1,\max}/W)H(x,M)+0.021/\rho d^2 - 1 > 0$. It is clear that the tunnel will automatically stable if $0.021/\rho d^2 > 1$ or $d < 0.145/\sqrt{\rho}$. For a sand quartz grain (density $2.6\times 10^3$ kg/m³) [10], we will always get the stability when $d< 3$ mm. It is also clear that, the grain size that support tunnel stability increases when the grain density decreases due to reduction in grain weight.

We inspect the effect of granular bond order on the maximum number of layers that support the tunnel stability, specifically for $Bo_g < 1$. We fix the ratio $zF_1/W = 50$ and we determine for three values of $x = 25$, 50, and 100. If we maintain the grain size, then different $x$ means different $R_1$ ($x = R_1/d$). Since the grain size is constant, the variation in $Bo_g$ merely was caused by variation in liquid bridge propeties (either the surface tension or parameter $\alpha$). **Figure 6(A)** is plot of maximum number of layer that maintaining the tunnel stability as function of $Bo_g$ for: (a) $x = 25$, (b) $x = 50$, and (c) $x = 100$. We express the vertical axis as $\ln(M/50)$. The numerical results have been well fitted with a general function $\ln(M/50) = \phi\exp(\gamma Bo_g)$, or

$$M = 50\exp[\phi\exp(\gamma Bo_g)] \tag{11}$$

with $\phi$ and $\gamma$ are parameters obtained from fitting process, the values of them are shown in **Table 1**. The three fitting processes all produced $R^2 = 0.9982$, indicated that the fitting results are very accurate. It is clear the maximum $M$ increases as $x$ decreases. As we have mentioned, since grain size is constant, the decrease in $x$ merely due to decrease in the tunnel radius. Therefore we conclude that for a specified grain size, the maximum height of the tunnel increases when the tunnel radius decreases. In is also interesting to see from Eq. (11) that the number 50 is likely a magic number, satisfied by all curves.

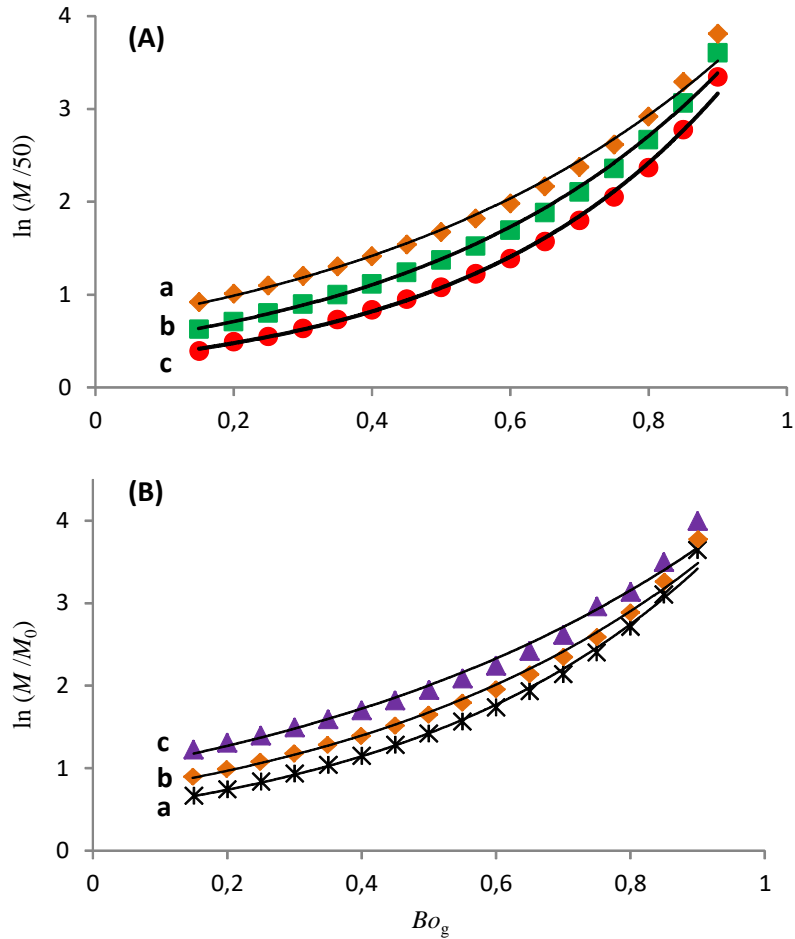

**Figure6**(A) The maximum number of layers that maintaining the tunnel stability as function of $Bo_g$ for: (a) $x = 25$, (b) $x = 50$, and (c) $x = 100$. We fix the ratio $zF_1/W = 50$. Symbols are numerical results and curves are fitting resultsusing a function expressed in Eq. (11) with the parameters are listed in Table 1. (B) The maximum number of layers that maintaining the tunnel stability as function of $Bo_g$ for: (a) $zF_1/W = 25$, (b) $x = 50$, and (c) $x = 100$. We fixed $x = 25$. Symbols are numerical results and curves are fitting results using a function $\ln(M/M_0) = \phi \exp(\gamma Bo_g)$ with the parameters are listed in **Table 1**.

**Table 1** Fitting parameters for curves in **Figure 6**.

| Data | $x$ | $zF_1/W$ | $\phi$ | $\gamma$ | $R^2$ |
|---|---|---|---|---|---|
| | **Parameters for Figure 6(A)** | | | | |
| a | 25 | 25 | 0.674 | 1.83 | 0.9982 |
| b | 50 | 25 | 0.441 | 2.29 | 0.9982 |
| c | 100 | 25 | 0.272 | 2.70 | 0.9982 |
| | **Parameters for Figure 6(B)** | | | | |
| | $M_0$ | $zF_1/W$ | $\phi$ | $\gamma$ | $R^2$ |
| a | 25 | 25 | 0.470 | 2.20 | 0.9981 |
| b | 50 | 50 | 0.674 | 1.83 | 0.9957 |
| c | 100 | 100 | 0.941 | 1.51 | 0.9910 |

**Figure 6(B)** is plot of maximum number of layer that maintaining the tunnel stability as function of $Bo_g$ for: (a) $zF_1/W = 25$, (b) $zF_1/W = 50$, and (c) $zF_1/W = 100$. We express the vertical axis as $\ln(M/M_0)$. The numerical results have been well fitted with a general function $\ln(M/M_0) = \phi \exp(\gamma Bo_g)$. Based on **Table 1**, it becomes clear that $M_0 = zF_1/W$. Then we obtain the general equation for the dependence of maximum $M$ to supprt the tunnel stability on other parameters as

$$M = \frac{zF_1}{W} \exp[\phi \exp(\gamma Bo_g)] \tag{12}$$

It is clear from Eq. (12), the number of layers of stable tunnel increases linearly with $zF_1/W$. If the suppoting force is absent ($F_1 = 0$) we are uanble to make tunnel. In addition, since $F_1$ has a maximum value $F_{1,max}$, we only able to made tunnel with maximum number of layers of $M_{max} = (zF_{1,max}/W)\exp[\phi \exp(\gamma Bo_g)]$, which is strongly dependent on the $Bo_g$. If $Bo_g = 0$, the maximum number of layers that can be made is $M_{max} = zF_{1,max}e^\phi/W$. For $zF_{1,max}/W \approx 100$ and based on **Table 2** that $\phi \approx 0.941$, the estimated maximum number of layers is 256 layers. For grain with diameter of 0.5 mm (the grain size used in this work), this number of layers corresponds to the tunnel thickness of 12.8 cm, which is nearly the same as the thickness of tunnel explored here of 14 cm.

It is also clear from Eq. (12) that, when $F_1 = 0$, $M = 0$ to mean that we are unable to make tunnel when the supporting force is zero. But we must remember that Eq. (12) applies for $Bo_g < 1$ only. Figure 6(A) and (B) are calculated results for $Bo_g < 1$. As shown by Eq. (7), we can always make a stable tunnel when $Bo_g > 1$, even for condition of $F_1 = 0$.

**CONCLUSION**

We have demonstrated the stability of tunnels made of granules. We identify the tunnel stability is strongly dependent on the grain size, tunnel diameter, and water content in the granules and density of the grain. Larger tunnel radius, larger grain size, and too much water content, and larged grain density tend to destabilize the tunnel. A phase diagram of stability was shown, exhibiting that the granular bond order of equal to 1 separated the region of totally stable and the region of partially stable. The region of totally stable means that the tunnel can always be made, independent of how large the supporting force. However, the partial stable state means that the stable tunnel can be made only up to a certain thickness which dependent on the supporting force and granular bond order. We also derived a specific formula relating the maximum thickness of stable tunnel and other parameters which be able to explain the observed data.

**REFERENCES**

[1] P. C. F. Møller and D. Bonn, EPL **80**, 38002 (2007).

[2] T.G. Mason, A. J. Levine, D. Ertas, and T.C. Halsey, Phys. Rev. E **60**, R5044-R5047 (1999).

[3] D.J. Hornbaker, R. Albert, I. Albert, A.-L.Barabi, and P. Schiffer, Nature **387**, 765 (1997).

[4] P. Schiffer, Nat. Phys. **1**, 21 (2005).

[5] S. Nowak, A. Samadani, and A. Kudrolli, Nat. Phys. **1**, 50 (2005).

[6] T.W. Lambe and R.V. Whitman, Soil Mechanics, New York: John Wiley & Sons (1969).

[7] T.C. Halsey and A.J. Levine, Phys. Rev. Lett. **80**, 3141 (1998).

[8] G. Levebvre and P. Jop, Phys. Rev. E **88**, 032205 (2013).

[9] S.R. Nagel, Rev. Mod. Phys. **64**, 321 (1992).

[10] A.Castellanos, Adv. Phys. 54, 263 (2005).

[11] N.D. Denkov, I.B. Ivanov, I. B. and P.A., Kralchevsky, J Colloid Interface Sci. **150**, 589 (1992).

[12] G. Lian, C. Thornton, and M.J. Adams, J Colloid Interface Sci. **161**, 138 (1993).

[13] T. Groger, U. Tuzun, and D. Heyes, Powder Technol. **133**, 203 (2003).

[14] T. Weigert and S. Ripperger, Part. Part. Sys. Charac. **16**, 238 (1999).

[15] C.D. Willett, M.J. Adams, S.A. Johnson, and J.P.K. Seville, Langmuir **16**, 9396 (2000).

[16] Y.I. Rabinovich, M.S. Esayanur, and B.M. Moudgil, Langmuir **21**, 10992 (2005).

[17] A.Gladkyy and R. Schwarze, arXiv:1403.7926v2 [cond-mat.soft] 12 Sep 2014.

[18] Z. Fournier, D. Geromichalos, S. Herminghaus, M.M. Kohonen, F. Mugele, M. Scheel, M. Schultz, B. Schultz, Ch. Schie, R. Seemann, and A. Skudelny, J. Phys.: Condens. Matter **17**, S477 (2005).